\documentclass[12pt]{article}
\usepackage[english]{babel}
\usepackage[latin1]{inputenc}
\usepackage[active]{srcltx}
\usepackage{amsfonts,amssymb,amsmath, epsfig}
\usepackage{color,graphicx,graphics,psfrag}
\usepackage{amsmath,amstext,amssymb,amsfonts, amscd}
\usepackage[dvipdfm]{hyperref}           

\def\Box{\leavevmode\vbox{\hrule
     \hbox{\vrule\kern4pt\vbox{\kern4pt}%
           \vrule}\hrule}}
\def\blackbox{\leavevmode\vrule height 5pt width 4pt depth 0pt\relax}
\def\endproof{\null\hfill {$\blackbox$}\bigskip}

\newcounter{appendix}
\def\appendix{\advance\c@appendix by 1
   \def\thesection{\Alph{section}}
   \ifnum\c@appendix=1 \setcounter{section}{-1} \fi
   \@startsection {section}{1}{\z@}{-3.5ex plus -1ex minus 
   -.2ex}{2.3ex plus .2ex}{\Large\bf}}

\def\paragraph#1{{\bf #1\ }}

\newtheorem{lemma}{Lemma}[section]  

\newtheorem{theorem}[lemma]{Theorem}

\newtheorem{proposition}[lemma]{Proposition}

\newtheorem{remark}{Remark}[section]

\title{Hydrodynamic models of self-organized dynamics: derivation and existence theory} 
\author{P. Degond $^{(1,2)}$, J-G. Liu$^{(3)}$, S. Motsch$^{(4)}$, V. Panferov$^{(5)}$} 
\date{} 
\begin{document}

\maketitle


\begin{center}
1-Université de Toulouse; UPS, INSA, UT1, UTM ;\\ 
Institut de Mathématiques de Toulouse ; \\
F-31062 Toulouse, France. \\
2-CNRS; Institut de Mathématiques de Toulouse UMR 5219 ;\\ 
F-31062 Toulouse, France.\\
email: pierre.degond@math.univ-toulouse.fr
\end{center}

\begin{center}
3- Department of Physics and Department of Mathematics\\
Duke University\\
Durham, NC 27708, USA\\
email: jliu@phy.duke.edu
\end{center}

\begin{center}
4-Department of Mathematics\\ 
University of Maryland\\
College Park, MD 20742-4015\\
email: smotsch@cscamm.umd.edu
\end{center} 

\begin{center}
5- Department of Mathematics \\
California State University, Northridge \\
18111 Nordhoff St \\
Northridge, CA 91330-8313 \\
email: vladislav.panferov@csun.edu

\end{center}

\begin{abstract}
  This paper is concerned with the derivation and analysis of hydrodynamic models for systems of self-propelled particles subject to alignment interaction and attraction-repulsion. The starting point is the kinetic model considered in \cite{DM} with the addition of an attraction-repulsion interaction potential. Introducing different scalings than in \cite{DM}, the non-local effects of the alignment and attraction-repulsion interactions can be kept in the hydrodynamic limit and result in extra pressure, viscosity terms and capillary force. The systems are shown to be symmetrizable hyperbolic systems with viscosity terms. A local-in-time existence result is proved in the 2D case for the viscous model and in the 3D case for the inviscid model. The proof relies on the energy method.
\end{abstract}

\medskip
\noindent
{\bf Acknowledgements:} The authors wish to acknowledge the hospitality of Mathematical Sciences Center and Mathematics Department of Tsinghua University where this research was performed. The research of J.-G. L. was partially supported by NSF grant DMS 10-11738.


\vskip 0.4cm

\setcounter{equation}{0}
\section{Introduction}
\label{sec_intro}

The context of this paper is the hydrodynamic limit of a kinetic model for self-propelled particles. The self-propulsion speed is supposed to be constant and identical for all the particles. Therefore, the velocity variable reduces to its orientation. 
The particle interactions consist in two parts: an alignment rule which tends to relax the particle velocity to the local average orientation and an attraction-repulsion rule which makes the particles move closer or farther away from each other. This model is inspired both by the Vicsek model \cite{Vicsek} and the Couzin model \cite{Aoki, Couzin}.

The model studied in this paper is a generalization of the model of \cite{DM} with the addition of an attraction-repulsion interaction potential. More importantly, a different scaling is investigated. In this scaling, the non-local effects of the alignment and attraction-repulsion interactions are kept in the hydrodynamic limit and result in extra pressure and viscosity terms. Beyond the statement of the model, the main result of the present paper is a local-in-time existence theorem in the 2D case for the viscous model (when the non-local effects are retained) and in the 3D case for the inviscid model (when the non-local effects are omitted). Both proofs rely on a suitable symmetrization of the system and on the energy method.

There has been an intense literature about the modeling of interactions between individuals among animal societies such as fish schools, bird flocks, herds of mammalians, etc. We refer e.g. to \cite{Aldana_Huepe, Aoki, Couzin, Gregoire_Chate} but an exhaustive bibliography is out of reach. Among these models, the Vicsek model \cite{Vicsek} has received particular attention due to its simplicity and the universality of its qualitative features. This model is a discrete particle model (or 'Individual-Based Model' or 'Agent-Based model') which consists of a time-discretized set of Ordinary Differential Equations for the particle positions and velocities. A time-continuous version of this model and its kinetic formulation are available in \cite{DM}.  A rigorous derivation of this kinetic model from the time-continuous Vicsek model can be found in \cite{BCC}. In the present paper, we extend this model by adding an attraction-repulsion force.

Hydrodynamic models are attractive over particle ones due to their computational efficiency. For this reason, many such models have been proposed in the literature \cite{CDP, CKMT, Chuang, Dorsogna, Mogilner1, Mogilner2, TB1, TB2}. However, most of them are phenomenological. \cite{DM} proposes one of the first rigorous derivations of a hydrodynamic version of the Vicsek model (see also \cite{KRZB,RBKZ,RKZB} for phenomenological derivations). It has been expanded in \cite{DM2} to account for a model of fish behavior where particles interact through curvature control, and in \cite{DY} to include diffusive corrections. Other variants have also been investigated. For instance, \cite{Frouvelle} studies the influence of a vision angle and of the dependency of the alignment frequency upon the local density. \cite{DFL,FL} propose a modification of the model which results in phase transitions from disordered to ordered equilibria as the density increases and reaches a threshold, in a way similar to polymer models \cite{DE, Onsager}.

The organization of the paper is as follows. In section \ref{sec:derivation}, we introduce the model of self-propelled particles and set up the associated kinetic equation. We then discuss various scalings which lead to the derivation of the studied hydrodynamic models. We introduce four dimensionless parameters in the problem: the scaled interaction mean-free path $\varepsilon$, the radius of the interaction region $\eta$, the noise intensity $\delta / \varepsilon$ and the relative strength between the attraction-repulsion and the alignment forces $s$. The scaling considered in \cite{DM} ignores the attraction-repulsion  force and supposes that $\varepsilon = \eta \to 0$, $\delta = O(1)$. Here, we investigate four different scaling relations. 
\begin{enumerate}
\item The weakly non-local interaction scaling without noise: $\eta = \sqrt \varepsilon$, $\delta = 0$, $s = \eta^2$. The resulting model is a viscous hydrodynamic model with constrained velocity on the unit sphere. For this case, we assume that the solutions of the kinetic equation are monokinetic. We justify this assumption by studying the space homogeneous kinetic model and prove that the solutions converge on the fast $\varepsilon$ time scale to the monokinetic distribution. We also highlight the variational structure of this space homogeneous kinetic model. Note that the scaling assumption $\eta = \sqrt \varepsilon$ is different from the one used in \cite{DM}. It corresponds to increasing the size of the interaction region in the microscopic variables by a factor $1/\sqrt \varepsilon$, as $\varepsilon \to 0$. Therefore, more and more non-local effects are picked up in the hydrodynamic limit. These non-local effects give rise to the viscosity terms in the macroscopic models which make an original addition from previous work. 
\item The local interaction scaling with noise. This is the scaling proposed in \cite{DM} which is recalled here just for the sake of comparisons. It consists in $\eta \ll \varepsilon$, $\delta = 0(1)$, $s \leq \eta^2$. The resulting model is the inviscid hydrodynamic model with constrained velocity on the unit sphere. 
\item The weakly non-local interaction scaling with noise. This scaling unifies the two previous scalings. It consists in $\eta = \sqrt \varepsilon$, $\delta = 0(1)$, $s = \eta^2$. Again, the resulting model is a viscous hydrodynamic model with constrained velocity on the unit sphere, but with modified coefficients as compared to the first scaling. We note however, that in the zero noise limit $\delta \to 0$, we recover the system obtained with the first scaling, which provides another justification of the monokinetic assumption in the derivation of the model. 
\item Capillary force scaling. This corresponds to $\eta = \sqrt \varepsilon$, $\delta = 0(1)$, $s = 1$.  Therefore, here, the attraction repulsion force is of the same order as the alignment force. However, we make the additional assumption that the zero-th order moment of the potential is zero, which expresses some kind of balance between the attraction and repulsion effects. This results in a model like in the previous scaling, but with the addition of a term analog to the capillary force, induced from the attractive part of the potential. 
\end{enumerate}

In section \ref{sec:existence_theory}, we prove local well-posedness for all the models derived in section \ref{sec:derivation}, except the last one (capillary force scaling). All the remaining systems have the same form of a symmetrizable hyperbolic system with additional viscosity. In section \ref{sub:existence_2D}, we prove the local-in-time existence of solutions for the viscous system in 2D and in section \ref{sub:existence_3D}, we show the same result for the inviscid system in 3D based on the energy method. Finally, a conclusion is drawn in section \ref{sec:conclu}.

\setcounter{equation}{0}
\section{Derivation of hydrodynamic models}
\label{sec:derivation}

\subsection{Individual-Based Model of self-alignment with attraction-repulsion}
\label{subsec:model}

\begin{figure}[ht]
  \centering
  \includegraphics[scale=1]{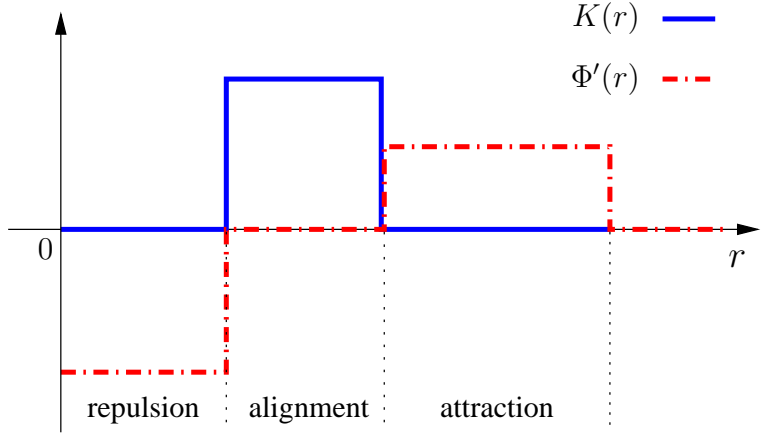}
  \caption{An example of functions $K$ and $\Phi'$ modeling the ``zone-based'' model: repulsion-alignment-attraction.}
  \label{fig:example_K_Phi}
\end{figure}

The starting point of this study is an Individual-Based Model of particles interacting through self-alignment \cite{Vicsek} and attraction-repulsion \cite{Aoki, Couzin}. Specifically, we consider $N$ particles $x_k\in\mathbb{R}^d$ moving at a constant speed $v_k \in\mathbb{S}^{d-1}$. Each particle adjusts its velocity to align with its neighbors and to get closer or further away. The evolution of each particle is modeled by the following dynamics:
\begin{eqnarray}
  \label{eq:particle_1}
  \frac{dx_k}{dt} &=& v_k \\
  dv_k &=& P_{v_k^\bot} \big(\overline{v}_k\,dt + \sqrt{2d}\; d\!B_t^k \big).
\end{eqnarray}
Here, $P_{v_k^\bot}$ is the projection matrix onto the normal plane to $v_k$:
\begin{displaymath}
  P_{v^\bot} = \mbox{Id} - v \otimes v.
\end{displaymath}
It ensures that $v_k$ stays of norm $1$. $B_t^k$ is a Brownian motion and $d$ represents the noise intensity. Both the alignment and attraction-repulsion rules are encoded in the vector $\overline{v}_k$:
\begin{displaymath}
  \overline{v}_k = \frac{j_k + r_k}{|j_k + r_k|},
\end{displaymath}
where $j_k$ counts for the alignment and $r_k$ for the attraction-repulsion:
\begin{equation}
  \label{eq:j_r_particle}
  j_k = \sum_j K(|x_j-x_k|) v_j \qquad,\qquad r_k = \sum_j \Phi'(|x_j-x_k|) \frac{x_j-x_k}{|x_j-x_k|}.
\end{equation}
The kernel $K$ is a positive function, $\Phi'$ can be both negative (repulsion) and positive (attraction). In figure \ref{fig:example_K_Phi}, we give an example of functions $K$ and $\Phi'$ modeling the popular ``zone-based'' model for fish behavior \cite{Aoki,Couzin,Parrish}.

When the number of particles becomes large (i.e. $N\rightarrow\infty$), one can formally derive the equation satisfied by the particle distribution function $f(x,v,t)$ (i.e. the probability distribution of the particles in phase-space $(x,v)$). Under suitable assumptions \cite{BCC,DM,sznitman_topics_1989}, $f$ satisfies:
\begin{eqnarray}
&& \hspace{-1cm} f_t + v \cdot \nabla_x f = - \nabla_v \cdot \left[ (P_{v^\bot} v_f) f \right] + d \Delta_v f,
\label{eq:kinetic_without_scaling}
\end{eqnarray}
where 
\begin{eqnarray*}
&& \hspace{-1cm}  v_f = \frac{j_f+r_f}{|j_f+r_f|},  \\
&& \hspace{-1cm}    j_f = \int_{x',v'}\!\!\! K(|x' - x|) v' \, f(x',v',t) \, dx'dv'  , \\ 
&& \hspace{-1cm} r_f = - \nabla_x \!\int_{x',v'}\!\!\!\Phi(|x' - x|) \, f(x',v',t) \, dx'dv'.
\end{eqnarray*}
The function $\Phi$ is the antiderivative of $\Phi'$ which vanishes at infinity (i.e. $\Phi(r) \stackrel{r \rightarrow \infty}{\longrightarrow} 0$). Using the distribution $f$, we want to identify the asymptotic behavior of the model in different regimes. This is the purpose of the next section.

\subsection{Scaling parameters}
\label{subsec_setting}

Introducing two dimensionless parameters $\varepsilon$ and $\eta$, the starting point is the following scaled version of the previous kinetic model for the distribution function $f(x,v,t)$:
\begin{eqnarray}
&& \hspace{-1cm} f_t + v \cdot \nabla_x f = - \frac{1}{\varepsilon} \nabla_v \cdot \left[ (P_{v^\bot} v_f^\eta) f \right] + d \Delta_v f,
\label{eq:kinetic_1}
\end{eqnarray}
where 
\begin{eqnarray} 
&& \hspace{-1cm} v_f^\eta = \frac{j_f^\eta + \eta^2 r_f^\eta}{|j_f^\eta + \eta^2 r_f^\eta|}, \label{eq:vfeta}\\
&& \hspace{-1cm} j_f^\eta =  \int_{(x',v') \in {\mathbb R}^n \times {\mathbb S}^{n-1}} K\left(\frac{|x' - x|}{\eta}\right) v' \, f(x',v',t) \, dx' \, dv' , \label{eq:jfeta} \\
&& \hspace{-1cm} r_f^\eta =  - \nabla_x \int_{(x',v') \in {\mathbb R}^n \times {\mathbb S}^{n-1}} \Phi\left(\frac{|x' - x|}{\eta}\right) \, f(x',v',t) \, dx' \, dv'. \label{eq:rfeta}
\end{eqnarray}
The first term (given by $j_f^\eta$) expresses the alignment interaction (like in the Vicsek dynamics \cite{Vicsek}) while the second term (given by $r_f^\eta$) expresses the repulsion interaction (like in e.g. \cite{Aoki}). The expression of $v_f^\eta$ means that the alignment term will prevail in the limit $\eta\rightarrow0$. In this paper, we will consider various possible assumptions concerning the relative speeds of convergence of $\varepsilon$ and $\eta$ to $0$. 

We denote:
\begin{eqnarray*}
&& \hspace{-1cm} \int_{\xi \in {\mathbb R}^{n}} K(|\xi|)  \, d\xi = k_0 , \\
&& \hspace{-1cm} \frac{1}{2n}  \int_{\xi \in {\mathbb R}^{n}} K(|\xi|) \, |\xi|^2 \, d\xi = k ,\\
&& \hspace{-1cm} \int_{\xi \in {\mathbb R}^{n}} \Phi (|\xi|)  \, d\xi = \phi . 
\end{eqnarray*}
We can always assume that $k_0=1$. The potential $\Phi$ is said to be repulsive if $\phi \geq 0$. Defining the moments $\rho_f$ and $\rho_f u_f$ of $f$ by 
\begin{eqnarray*}
&& \hspace{-1cm} \rho_f =  \int_{v' \in {\mathbb S}^{n-1}} f(v') \, dv', \\
&& \hspace{-1cm} \rho_f u_f  =  \int_{v' \in {\mathbb S}^{n-1}} f(v') \, v' \, dv', 
\end{eqnarray*}
we have the following Taylor expansion of $v_f^\eta$: 
\begin{eqnarray*}
&& \hspace{-1cm}  v_f^\eta = \Omega_f + \eta^2 \frac{1}{\rho_f |u_f|} \ell_f + o(\eta^2), \\
&& \hspace{-1cm}  \Omega_f = \frac{u_f}{|u_f|}, \quad \ell_f = P_{\Omega_f^\bot} ( k \Delta (\rho_f u_f) - \phi \nabla_x \rho_f ) . 
\end{eqnarray*}
Inserting this expression into the kinetic equation (\ref{eq:kinetic_1}), we get
\begin{eqnarray}
&& \hspace{-1cm} f_t + v \cdot \nabla_x f = - \frac{1}{\varepsilon} \nabla_v \cdot \left[ ( P_{v^\bot} \Omega_f ) f \right]  \nonumber \\
& & \hspace{1cm} - \frac{\eta^2}{\varepsilon} \frac{1}{\rho_f |u_f|} \nabla_v \cdot \left[ ( P_{v^\bot} \ell_f ) f \right] + d \Delta_v f + o(\frac{\eta^2}{\varepsilon})  . 
\label{eq:kinetic_2}
\end{eqnarray}

We now consider three different scaling limits which lead to models for which we will prove local existence of classical solutions.

\subsection{Weakly non-local interaction scaling without noise}
\label{sub:weak}

In this scaling limit, we assume no noise $d=0$ and the following ordering between the two parameters $\varepsilon$ and $\eta$: 
$$ \varepsilon \to 0, \quad \eta \to 0, \quad \frac{\eta^2}{\varepsilon} \to 1.$$
$f^\varepsilon$ satisfies (keeping only the $O(1)$ terms in $\varepsilon$ or larger): 
\begin{eqnarray}
&& \hspace{-1cm} f^\varepsilon_t + v \cdot \nabla_x f^\varepsilon = - \frac{1}{\varepsilon} \nabla_v \cdot \left[ ( P_{v^\bot} \Omega_{f^\varepsilon} ) f^\varepsilon \right]  -  \frac{1}{\rho_{f^\varepsilon} |u_{f^\varepsilon}|} \nabla_v \cdot \left[ ( P_{v^\bot} \ell_{f^\varepsilon} ) f^\varepsilon \right]   . 
\label{eq:kinetic_2.1}
\end{eqnarray}

For $f^\varepsilon$  to converge, we need to assume that the leading order term at the right-hand side of (\ref{eq:kinetic_2.1}) vanishes, i.e. that  $f^\varepsilon$ satisfies:  
$$\nabla_v \cdot \left[ ( P_{v^\bot} \Omega_{f^\varepsilon} ) f^\varepsilon \right] = 0, \quad \forall \varepsilon > 0.$$
This is equivalent to assuming that $f^\varepsilon$ is a monokinetic distribution, i.e. 
\begin{equation}
f^\varepsilon (x,v,t) = \rho^\varepsilon(x,t) \, \delta(v,\Omega^\varepsilon(x,t)),
\label{eq:monokinetic}
\end{equation}
where $\delta(v,\bar v)$ is the delta distribution on the sphere at the point $\bar v$. 
The assumption of monokinetic distribution requires, in order to be consistent, that there is no noise, which is the reason for assuming $d=0$.

\begin{proposition}
For monokinetic solutions (\ref{eq:monokinetic}), $\rho$ and $\Omega$ are independent of $\varepsilon$ and satisfy the following system: 
\begin{eqnarray}
&& \hspace{-1cm} \partial_t \rho + \nabla_x \cdot (\rho \Omega) = 0 , \label{noiseless_mass} \\
&& \hspace{-1cm} \partial_t (\rho \Omega) + \nabla_x \cdot (\rho \Omega \otimes \Omega) + \phi P_{\Omega^\bot} \nabla_x \rho = k P_{u^\bot} \Delta (\rho u) . \label{noiseless_momentum} 
\end{eqnarray}
\label{prop:scaling_1}
\end{proposition}

\medskip
\noindent
{\bf Proof.} The result follows from  multiplying (\ref{eq:kinetic_2.1}) by $1$ and $v$ and using Green's formula. \endproof

\begin{remark} 
The repulsive force contributes for a pressure term at the left-hand side of the momentum equation, which otherwise would not be strictly hyperbolic, and would fall in the class of Pressureless Gas Dynamics models \cite{Bouchut}.
\end{remark}

In order to justify the monokinetic assumption, we consider the spatial homogeneous equation:
\begin{eqnarray}
&& \hspace{-1cm} f_t = - \frac{1}{\varepsilon} \nabla_v \cdot \left[ ( P_{v^\bot} \Omega_f ) f \right]  , 
\label{eq:kinetic_4}
\end{eqnarray}
and show that its solution relaxes to a monokinetic distribution (\ref{eq:monokinetic}) at the fast $\varepsilon$ time scale. More precisely, we have

\begin{proposition}
We assume that $\int f|_{t=0} \, dv = 1$ so that we have $\int f(v,t) \, dv = 1$ for all times. Therefore, $u = \int v \, f(v) \, dv$ (we omit the index $f$ when the context is clear). We also assume that $u_{t=0} \not = 0$, otherwise, the dynamics is not defined. Then, $f(t) \to f(\infty)$ where $f(\infty)$ is of the form $f(\infty) = \delta(v,\Omega)$. 
\label{prop:monokinetic_convergence}
\end{proposition}

\medskip
\noindent
{\bf Proof.} We introduce the variance:
\begin{eqnarray*} {\mathcal F}(f) & = & \int |v-u|^2 \, f(v) \, dv  \\
& = & \int (1 - (u \cdot v)) \, f(v) \, dv .
\end{eqnarray*}
We have ${\mathcal F}(f) = 1 - |u|^2$. We note that $|u|^2$ is the classical order parameter \cite{Vicsek}. 
In Lemma \ref{lem:variance} below, we prove that ${\mathcal F}(f)$ satisfies the following dissipation equation:  
\begin{eqnarray}
&& \hspace{-1cm} {\mathcal F}(f)_t +  \frac{2}{\varepsilon}  \int \frac{|u|^2 - (v\cdot u)^2}{|u|} \, f(v) \, dv = 0. 
\label{eq:variance}
\end{eqnarray}
Since $|u|^2 - (v\cdot u)^2 \geq 0$, we have 
$$ {\mathcal F}(f)_t \leq 0 . $$
Therefore, ${\mathcal F}(f)$ is a decreasing function of time. Furthermore, if $f$ is a distribution such that 
$$\int \frac{|u|^2 - (v\cdot u)^2}{|u|} \, f(v) \, dv = 0, $$
then, either $f = \delta(v,\Omega)$ or $f$ is of the form:
$$ f = \alpha \delta(v,\Omega) + (1-\alpha) \delta(v,-\Omega) ,$$
where $\alpha \in [0,1]$ and $\Omega \in {\mathbb S}^{n-1}$. The second form is called a dipole. So, unless $f$ is a delta or a dipole, ${\mathcal F}(f)$ is strictly decaying. Since a dipole is unstable, it is never reached in the course of the dynamics. Therefore, ${\mathcal F}(f) \to 0$ as $t \to \infty$. Since ${\mathcal F}(f) = 0$ is equivalent to $f = \delta(v,\Omega)$, this shows that $f \to \delta(v,\Omega)$ as $t \to \infty$ and the typical convergence time is $\varepsilon$. \endproof

\begin{lemma}
Any solution $f$ of (\ref{eq:kinetic_4}) satisfies (\ref{eq:variance}).
\label{lem:variance}
\end{lemma}

\medskip
\noindent
{\bf Proof.} We write
\begin{eqnarray*}
\varepsilon {\mathcal F}(f)_t &=& \int (1 - (u \cdot v)) \, \partial_t f(v) \, dv - u_t \cdot \int v \, f(v) \, dv \\
&=&  \int (1 - (u \cdot v)) \, \partial_t f(v) \, dv - \int v \, \partial_t f(v) \, dv \cdot u \\
&=& \partial_t (\int f(v) \, dv ) - 2 \int (u \cdot v) \, \partial_t f(v) \, dv \\
&=& - 2 \int (u \cdot v) \, \partial_t f(v) \, dv .
\end{eqnarray*}
In the last equality, we have used that  $\int f(v,t) \, dv = 1$. 
Now, multiplying (\ref{eq:kinetic_4}) by $- 2 (u \cdot v)$, integrating with respect to $v$ and using Green's formula, we get:
\begin{eqnarray*}
&& \hspace{-1cm} \varepsilon {\mathcal F}(f)_t + 2 \int ( u \cdot P_{v^\bot} \Omega ) \,  f(v) \, dv =0 .
\end{eqnarray*}
But 
$$ u \cdot P_{v^\bot} \Omega = \frac{1}{|u|} (|u|^2 - (u \cdot v)^2), $$
which leads to the result. \endproof

${\mathcal F}(f)$ is a free energy for the problem (\ref{eq:kinetic_4}) and provides a variational structure. 
First, let us denote by $\nu = \nabla_f {\mathcal F}$ the gradient of ${\mathcal F}$ with respect to $f$. 
It is defined by
$$ \langle \nabla_f {\mathcal F} , g \rangle = \frac{\delta {\mathcal F}}{\delta f} (g) $$
where $g$ is an increment of $f$, i.e. a function $g(v)$ satisfying $\int g(v) \, dv = 0$ (so that $f + g$ satisfies the admissibility condition $\int (f+g) \, dv = 1$). 

\begin{proposition}
Eq. (\ref{eq:kinetic_4}) can be recast as 
\begin{eqnarray}
&& \hspace{-1cm} f_t - \frac{1}{\varepsilon |u|} \nabla_v \cdot \{ [\nabla_v  (\nabla_f {\mathcal F}) ] f\}  =0 , \label{eq:variational}
\end{eqnarray}
which shows that the flow of (\ref{eq:kinetic_4}) has a gradient flow structure in the Wasserstein metrics \cite{Villani}. We have:
\begin{eqnarray}
&& \hspace{-1cm} {\mathcal F}(f)_t + \frac{1}{\varepsilon |u|} \int |\nabla_v  (\nabla_f {\mathcal F}) |^2 \, f \, dv  =0 . \label{eq:variational_estimate}
\end{eqnarray}
\label{prop:gradient_structure}
\end{proposition}

\begin{remark}
Eq. (\ref{eq:variational_estimate}) provides another proof of the decay of ${\mathcal F}(f)$ with time. 
\end{remark}

\medskip
\noindent
{\bf Proof.} We compute: 
\begin{eqnarray*}
&& \hspace{-1cm} \frac{\delta {\mathcal F}}{\delta f} (g) =  - 2 u \cdot \int g(v) \, v \,  dv = \langle - 2 (u \cdot v) , g \rangle , 
\end{eqnarray*}
which yields
$$ \nu = \nabla_f {\mathcal F} = - 2 u \cdot v . $$
A simple computation shows that 
$$ \nabla_v \nu = - P_{v^\bot} u = - |u| P_{v^\bot} \Omega  . $$
Therefore, eq. (\ref{eq:kinetic_4}) can be written as (\ref{eq:variational}). Now, multiplying by $\nabla_f {\mathcal F}$, integrating over $v$ and using Green's formula, we get  (\ref{eq:variational_estimate}), which ends the proof. 
\endproof

\subsection{Local interaction scaling with noise}
\label{sub:local}

In this scaling we assume that $d = O(\frac{1}{\varepsilon})$. More precisely, we let: 
$$ d = \frac{\delta}{\varepsilon}, $$
with $\delta$ a given constant. 
We also assume that $\varepsilon$ and $\eta$ are such that: 
$$\varepsilon \to 0, \quad \eta \to 0, \quad \frac{\eta^2}{\varepsilon} \to 0.$$
With this last assumption, the $O(\frac{\eta^2}{\varepsilon})$ term in (\ref{eq:kinetic_2}), which results from the non-locality of the average alignment direction, vanishes. Therefore, this scaling keeps only the local contribution of the alignment interaction. 
The resulting asymptotic problem, keeping only terms of order $O(1)$ or larger, is written:
\begin{eqnarray}
&& \hspace{-1cm} f^\varepsilon_t + v \cdot \nabla_x f^\varepsilon = - \frac{1}{\varepsilon} \left\{ \nabla_v \cdot \left[ ( P_{v^\bot} \Omega_{^\varepsilon} ) f^\varepsilon \right] + \delta \Delta_v f^\varepsilon \right\}.
\label{eq:kinetic_3}
\end{eqnarray}
The limit of (\ref{eq:kinetic_3}) as $\varepsilon \to 0$ has been studied in \cite{DM} in dimension $3$ and in \cite{Frouvelle} in any dimensions. The result is stated in the following theorem. 

\begin{theorem}
  We have $f^\varepsilon \to \rho M_{\Omega}$ where $M_{\Omega}(v)$ is the Von Mises-Fischer distribution: 
  \begin{equation}
    M_{\Omega}(v) = \frac{\exp( \beta (v \cdot \Omega)) \, dv }{\int_{v \in {\mathbb S}^{n-1}} \exp( \beta (v \cdot \Omega)) \, dv}, \quad \beta = \frac{1}{\delta} , 
    \label{eq:VMF}
  \end{equation}
  and $\rho$ and $\Omega$ satisfy the following system:
  \begin{eqnarray}
    && \hspace{-1cm} \partial_t \rho + c_1 \nabla_x \cdot (\rho \Omega) = 0 , \label{DM_mass} \\
    && \hspace{-1cm} \rho (\partial_t \Omega + c_2 \Omega \cdot \nabla_x \Omega ) + \delta P_{\Omega^\bot} \nabla_x \rho = 0 . \label{DM_momentum}
  \end{eqnarray}
  The constants $c_1$ and $c_2$ are defined by 
  \begin{eqnarray*}
    && \hspace{-1cm} c_1 = \int_{v \in {\mathbb S}^{n-1}} M_{\Omega}(v)\, (v \cdot \Omega) \,  dv, \\
    && \hspace{-1cm} c_2 = \frac{\int_{v \in {\mathbb S}^{n-1}} M_{\Omega}(v) \, h (v \cdot \Omega) \, (1 - (v \cdot \Omega)^2) \,  (v \cdot \Omega) \,  dv}{\int_{v \in {\mathbb S}^{n-1}} M_{\Omega}(v) \, h (v \cdot \Omega) \, (1 - (v \cdot \Omega)^2) \, dv} ,
  \end{eqnarray*}
  where $h (v \cdot \Omega)$ is the Generalized Collision Invariants (GCI) \cite{DM} and is defined as follows in the $n$-dimensional case \cite{Frouvelle}. Set $\psi_a(v) = h(\Omega \cdot v) \,  (a \cdot v)$ where $a \in {\mathbb R}^n$ is any vector such that $a \cdot \Omega = 0$. Then, $\psi_a$ is the unique solution in the Sobolev space $H^1({\mathbb S}^{n-1})$ with zero mean, of the following elliptic problem: 
  \begin{eqnarray*}
    && \hspace{-1cm} - \Delta_v \psi - \beta (\Omega \cdot \nabla_v) \psi = a \cdot v.
  \end{eqnarray*}
  \label{thm:scaling_2}
\end{theorem}
\noindent {\bf Proof.} We refer to \cite{DM} in the three dimensional case and \cite{Frouvelle} in the general $n$-dimensional case. \endproof

\subsection{Weakly non-local interaction scaling with noise}
\label{sub:non-local}

In this section, we propose a scaling which unifies the two previous ones. In this scaling we assume that $d =  \frac{\delta}{\varepsilon}$, with $\delta$ is a given $O(1)$ constant and that: 
$$\varepsilon \to 0, \quad \eta \to 0, \quad \frac{\eta^2}{\varepsilon} \to 1.$$
Here $\frac{\eta^2}{\varepsilon} \to 1$ instead of $0$ like in the previous section. 
Inserting these assumptions into (\ref{eq:kinetic_2}), and keeping terms of order $O(1)$ or larger, we get
\begin{eqnarray}
&& \hspace{-1cm} f^\varepsilon_t + v \cdot \nabla_x f^\varepsilon + \frac{1}{\rho_{f^\varepsilon} |u_{f^\varepsilon}|} \nabla_v \cdot \left[ ( P_{v^\bot} \ell_{f^\varepsilon} ) f^\varepsilon \right]= \nonumber \\
& & \hspace{4cm} - \frac{1}{\varepsilon} \left\{ \nabla_v \cdot \left[ ( P_{v^\bot} \Omega_{f^\varepsilon} ) f^\varepsilon \right]  
 + \delta \Delta_v f^\varepsilon \right\}  . 
\label{eq:kinetic_6}
\end{eqnarray}
The limit $\varepsilon \to 0$ is investigated in the following theorem: 

\begin{theorem}
We have $f^\varepsilon \to \rho M_{\Omega}$ where $M_{\Omega}(v)$ is the Von Mises-Fischer distribution (\ref{eq:VMF}). $\rho$ and $\Omega$ satisfy the following system:
\begin{eqnarray}
&& \hspace{-1cm} \partial_t \rho + c_1 \nabla_x \cdot (\rho \Omega) = 0 , \label{New_mass} \\
&& \hspace{-1cm} \rho (\partial_t \Omega + c_2 \Omega \cdot \nabla_x \Omega ) + (\delta + c_3 \phi) P_{\Omega^\bot} \nabla_x \rho = c_3 k c_1 P_{\Omega^\bot} \Delta (\rho \Omega) , \label{New_momentum_2}
\end{eqnarray}
where the constants $c_1$ and $c_2$ are defined as in Theorem \ref{thm:scaling_2} and 
\begin{equation} 
c_3 = \frac{(n-1) \delta + c_2}{c_1}  . 
\label{eq:c3}
\end{equation}
\label{thm:scaling_3}
\end{theorem}

\begin{remark}
We notice that $c_k \to 1$ as $\delta \to 0$ for $k=1, 2, 3$ and we recover the noiseless system 
(\ref{noiseless_mass}), (\ref{noiseless_momentum}) when $\delta \to 0$. 
\end{remark}

\medskip
\noindent
{\bf Proof.} We write (\ref{eq:kinetic_6}) as 
$$ (T_1 + T_2) f^\varepsilon = \frac{1}{\varepsilon} Q(f^\varepsilon), $$
where $T_1 + T_2$ and $Q$ are respectively the operators appearing at the left and right hand sides of (\ref{eq:kinetic_6}). $T_1 = \partial_t + v \cdot \nabla_x$ and $T_2$ is the remaining part of the left-hand side. Integrating over $v$ and letting $f^\varepsilon \to  \rho M_{\Omega}$ leads to the mass conservation equation (\ref{DM_mass}) unchanged, since $T_2$ is in divergence form and vanishes through integration with respect to $v$. 

Now, to get the momentum equation, we proceed like in \cite{DM}. From the Generalized Collision Invariant property \cite{DM}, it follows that 
$$  P_{\Omega^\bot} \int_{v \in {\mathbb S}^{n-1}} T (\rho M_\Omega) \, h \, v \, dv = 0 .$$
Now, the term 
$$  P_1 := P_{\Omega^\bot} \int_{v \in {\mathbb S}^{n-1}} T_1 (\rho M_\Omega) \, h \, v \, dv,$$
gives rise to the same expression as in \cite{DM}. This expression is
$$ P_1 = \beta \alpha \rho \partial_t \Omega + \gamma \Omega \cdot \nabla_x \Omega  + \alpha P_{\Omega^\bot} \nabla_x \rho , $$
with 
\begin{eqnarray*}
&& \hspace{-1cm} \alpha  = \frac{1}{n-1} \int_{v \in {\mathbb S}^{n-1}} M_\Omega(v) \, h \, (1 - (v \cdot \Omega)^2) \, dv
, \\
&& \hspace{-1cm} \gamma = \frac{1}{(n-1) \delta} \int_{v \in {\mathbb S}^{n-1}} M_\Omega(v) \, h \, (1 - (v \cdot \Omega)^2) \,  (v \cdot \Omega) \,  dv.
\end{eqnarray*}
Dividing by $\alpha \beta$, we find the coefficients $c_2$ and $\delta$ of (\ref{DM_momentum}) (we recall that $\beta \delta = 1$). 

We introduce the notation 
$$ \ell := \ell_{\rho M_\Omega} = P_{\Omega^\bot} ( k c_1 \Delta (\rho \Omega) - \phi \nabla_x \rho ), $$
and consider 
\begin{eqnarray*}
P_2 &:=& P_{\Omega^\bot} \int_{v \in {\mathbb S}^{n-1}} T_2 (\rho M_\Omega) \, h \, v \, dv \\
&=& \frac{1}{c_1} P_{\Omega^\bot} \int_{v \in {\mathbb S}^{n-1}}  \nabla_v \cdot \left[ ( P_{v^\bot} \ell ) M_\Omega \right] \, h \, v \, dv .
\end{eqnarray*}
Using Green's formula, we get
\begin{eqnarray*}
P_2 &=& -\frac{1}{c_1} P_{\Omega^\bot} \int_{v \in {\mathbb S}^{n-1}} \left[  P_{v^\bot} \ell  \right] \cdot \nabla_v ( h \, v) \, M_\Omega\, dv.
\end{eqnarray*}
We note that $(P_{v^\bot} \ell) \cdot \nabla_v \phi = \ell \cdot \nabla_v \phi$, with $\phi$ being any component of $h \, v$. We deduce that 
\begin{eqnarray}
P_2 &=& -\frac{1}{c_1} P_{\Omega^\bot} \int_{v \in {\mathbb S}^{n-1}}  (\ell \cdot \nabla_v) ( h \, v) \, M_\Omega\, dv  \nonumber \\
&=& -\frac{1}{c_1} P_{\Omega^\bot} \left( \int_{v \in {\mathbb S}^{n-1}} \nabla_v ( h \, v) \, M_\Omega\, dv \right)^T \ell . \label{eq:P2}
\end{eqnarray}
Now, we use the formulas: 
\begin{eqnarray*} 
& & \int_{{\mathbb S}^{n-1}} \nabla_v g \, dv = (n-1) \int_{{\mathbb S}^2} v g \, dv \\
& & \int_{{\mathbb S}^{n-1}} (\nabla_v g) h \, dv = (n-1) \int_{{\mathbb S}^2} v g h \, dv -  \int_{{\mathbb S}^2} (\nabla_v h) g \, dv
\end{eqnarray*}
for any pair of scalar functions $g$, $h$ on ${\mathbb S}^{n-1}$. We recall that $\nabla_v M_\Omega = \beta P_{v^\bot} \Omega M_\Omega$. Since $\ell \cdot \Omega = 0$, we compute the matrix 
\begin{eqnarray*}
D &:=&  \left( \int_{v \in {\mathbb S}^{n-1}} \nabla_v ( h \, v) \, M_\Omega\, dv \right)^T P_{\Omega^\bot}  \\ 
&=& (n-1)  \left( \int_{v \in {\mathbb S}^{n-1}} (v \otimes v) h \, M_\Omega\, dv \right)  P_{\Omega^\bot} \\
& & \hspace{4cm} -  \beta  \left( \int_{v \in {\mathbb S}^{n-1}} (v \otimes P_{v^\bot} \Omega)  h \, M_\Omega\, dv \right) P_{\Omega^\bot} \\
&:=& (n-1) D_1 -  D_2.
\end{eqnarray*}
We decompose 
$$ v = v_\bot + v_\parallel, \quad v_\bot = P_{\Omega^\bot} v, \quad v_\parallel = (v \cdot \Omega) \Omega. $$
Using this decomposition and the fact that integrals of odd degree polynomials of $v_\bot$ over ${\mathbb S}^{n-1}$ vanish, we have: 
\begin{eqnarray*}
D_1 &=&    \left( \int_{v \in {\mathbb S}^{n-1}} (v_\bot \otimes v_\bot) h \, M_\Omega\, dv \right)  P_{\Omega^\bot} = \alpha P_{\Omega^\bot} ,
\end{eqnarray*}
and
\begin{eqnarray*}
D_2 &=&  \beta  \left( \int_{v \in {\mathbb S}^{n-1}} ((v_\parallel + v_\bot ) \otimes (\Omega - (\Omega \cdot v)) (v_\parallel + v_\bot)) h \, M_\Omega\, dv \right)  P_{\Omega^\bot} .
\end{eqnarray*}
Owing to the fact that any term of the form $(A \otimes v_\parallel)  P_{\Omega^\bot} = 0$ for any vector $A$, we have since $v_\parallel$ is parallel to $\Omega$:
\begin{eqnarray*}
D_2 &=&   - \beta \left( \int_{v \in {\mathbb S}^{n-1}} (v_\bot \otimes v_\bot) (\Omega \cdot v) \,  h \, M_\Omega\, dv \right)  P_{\Omega^\bot}  = - \gamma P_{\Omega^\bot}.
\end{eqnarray*}
Inserting these results into (\ref{eq:P2}), we get
\begin{eqnarray*}
P_2 &=& -\frac{(n-1) \alpha + \gamma}{c_1} \ell .
\end{eqnarray*}
Collecting all the results and dividing by $\alpha \beta$, we are led to the momentum equation  (\ref{New_momentum_2}), which ends the proof. \endproof

\subsection{Attraction-repulsion potential: induced capillary force}
\label{sub:capillary}

In this section, we investigate the case where the attraction-repulsion force term is of the same order as the alignment term in the expression of the alignment direction $v_f^\eta$, i.e. we assume that 
\begin{eqnarray} 
&& \hspace{-1cm} v_f^\eta = \frac{j_f^\eta +  r_f^\eta}{|j_f^\eta +  r_f^\eta|}, \label{eq:vfeta2}
\end{eqnarray}
where $j_f^\eta$ and $r_f^\eta$ are respectively given by (\ref{eq:jfeta}) and (\ref{eq:rfeta}). Note that, by contrast to (\ref{eq:vfeta}), there is no $\eta^2$ in front of $r_f^\eta$ in (\ref{eq:vfeta2}). 

The Taylor expansion of $v_f^\eta$ is now given by 
\begin{eqnarray*}
&& \hspace{-1cm}  v_f^\eta = \hat \Omega_f + \eta^2 \frac{1}{\rho_f |u_f|} \ell_f + o(\eta^2), \\
&& \hspace{-1cm}  \hat \Omega_f = \frac{u_f - \phi \nabla_x \rho_f }{|u_f - \phi \nabla_x \rho_f|}, \quad \ell_f = P_{\hat \Omega_f^\bot} ( k \Delta (\rho_f u_f) - \phi_2 \nabla_x \Delta \rho_f ),
\end{eqnarray*}
where 
\begin{eqnarray*}
&& \hspace{-1cm} \frac{1}{2n} \int_{x' \in {\mathbb R}^{n}} \Phi (|\xi|)  \, |\xi|^2  \, d\xi = \phi_2 . 
\end{eqnarray*}

\begin{figure}
\begin{center}
\input{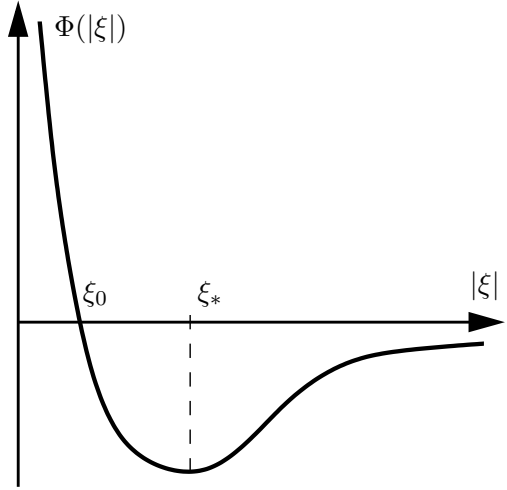}
\caption{The attraction-repulsion potential $\Phi$}
\label{fig:potential}
\end{center}
\end{figure}

Here, we suppose like in \cite{Aoki}, that the potential is repulsive at short scales and attractive at large scales (See Fig. \ref{fig:potential}). Therefore, $ \Phi (|\xi|)$ is supposed to decrease for  $|\xi| \in [0, \xi_*]$ and to increase for $|\xi| \in [\xi_*, + \infty)$. Furthermore, since $\Phi (|\xi|) $ is supposed integrable on ${\mathbb R}^n$, we have $\Phi(|\xi|) \to 0$ as $|\xi| \to \infty$. It results that $\Phi(\xi_*) < 0$ and that $\Phi \geq 0 $ for  $|\xi| \in [0, \xi_0]$ and $\Phi \leq 0 $ for $|\xi| \in [\xi_0, + \infty)$ where $\xi_0 < \xi_*$. We make the additional assumption that the zero-th order moment vanishes:
$$\phi = 0, $$
which expresses the balance between the attractive and repulsive parts of $\Phi$. Given the above assumptions, the second moment is negative: 
$$\phi_2 < 0. $$  

With these assumptions, the Taylor expansion of $v_f^\eta$ simplifies and becomes:
\begin{eqnarray}
&& \hspace{-1cm}  v_f^\eta = \Omega_f + \eta^2 \frac{1}{\rho_f |u_f|} \ell_f + o(\eta^2), \nonumber \\
&& \hspace{-1cm}  \Omega_f = \frac{u_f}{|u_f|}, \quad \ell_f = P_{\Omega_f^\bot} ( k \Delta (\rho_f u_f) - \phi_2 \nabla_x \Delta \rho_f ) . \label{eq:ell_new}
\end{eqnarray}
Now, we can develop the same theory as before, assuming that 
$$\varepsilon \to 0, \quad \eta \to 0, \quad \frac{\eta^2}{\varepsilon} \to 1.$$
Inserting these assumptions into (\ref{eq:kinetic_2}), and keeping terms of order $O(1)$ or larger, we get
(\ref{eq:kinetic_6}) but with $\ell$ given by (\ref{eq:ell_new}). The limit $\varepsilon \to 0$ can be performed like in section \ref{sub:non-local} and we obtain the following theorem:

\begin{theorem}
We have $f^\varepsilon \to \rho M_{\Omega}$ where $M_{\Omega}(v)$ is the Von Mises-Fischer distribution (\ref{eq:VMF}). $\rho$ and $\Omega$ satisfy the following system:
\begin{eqnarray}
&& \hspace{-1cm} \partial_t \rho + c_1 \nabla_x \cdot (\rho \Omega) = 0 , \label{Newnew_mass} \\
&& \hspace{-1cm} \rho (\partial_t \Omega + c_2 \Omega \cdot \nabla_x \Omega ) + \delta P_{\Omega^\bot} \nabla_x \rho = c_3 k c_1 P_{\Omega^\bot} \Delta (\rho \Omega) + c_3 |\phi_2| P_{\Omega^\bot} \nabla_x \Delta \rho, \label{Newnew_momentum_2}
\end{eqnarray}
where the constants $c_1$ and $c_2$ are defined as in Theorem \ref{thm:scaling_2} and $c_3$ as in Theorem \ref{thm:scaling_3}.
\label{thm:scaling_4}
\end{theorem}

\begin{remark}
The last term at the right-hand side of (\ref{Newnew_momentum_2}) has the same expression as the capillary force in fluid dynamics, except for the projection operator $P_{\Omega^\bot}$. This capillary force is induced from the attractive part of the potential $\Phi$.
\end{remark}

\setcounter{equation}{0}
\section{Existence theory}
\label{sec:existence_theory}

\subsection{Existence in 2D with viscosity}
\label{sub:existence_2D}

This section is concerned with a local existence result in 2D for a system of the general form
\begin{eqnarray}
&& \hspace{-1cm} \partial_t \rho + \nabla_x \cdot (\rho \Omega) = 0 , \label{Gene_mass} \\
&& \hspace{-1cm} \rho (\partial_t \Omega + c \Omega \cdot \nabla_x \Omega ) + P_{\Omega^\bot} \nabla_x (p(\rho)) = \mu P_{\Omega^\bot} \Delta (\rho \Omega) , \label{Gene_momentum}
\end{eqnarray}
where the constants $c \in {\mathbb R}$ and $\mu \geq 0$ are given and the pressure relation $p(\rho)$ satisfies $p'(\rho) >0$. All systems derived in the previous section can by recast in this form, with a particular choice of $c$, $\mu$ and $p(\rho)$, after time rescaling, except for the last one (section \ref{sub:capillary}) involving the capillary force. The system is supplemented with initial data $\rho_0 > 0$ and $\Omega_0$ such that $|\Omega_0| = 1$. We assume that the domain is the square box $\Pi^2= [0,1]^2$ with periodic boundary conditions.  

\begin{theorem}
We assume that the initial data belong to $H^m(\Pi^2)$ with $m>2$. Then, there exists a time $T>0$ and a unique solution $(\rho, \varphi)$ in $L^\infty([0,T],H^m(\Pi^2)) \cap H^1([0,T],$ $H^{m-1}(\Pi^2))$ such that $\rho$ remains positive. If, in addition, $\mu>0$, then, the solution also belongs to $L^2([0,T],$ $H^{m+1}(\Pi^2))$. 
\label{thm:exist_2D}
\end{theorem}

\medskip
\noindent
{\bf Proof. } In 2D, we can set $\Omega = (\cos \varphi, \sin \varphi)$. We recall that 
$$ \partial_t \Omega = \Omega^\bot \, \partial_t \varphi, \quad \nabla_x \cdot \Omega = (\Omega^\bot \cdot \nabla_x) \varphi, \quad P_{\Omega^\bot} = \Omega^\bot \otimes \Omega^\bot, $$
with $\Omega^\bot = (- \sin \varphi, \cos \varphi)$. Then, we have
\begin{eqnarray*}
& & \hspace{-1cm} \Delta (\rho \Omega) = \Delta \rho \, \,  \Omega + 2 \Omega^\bot (\nabla_x \rho \cdot \nabla_x \varphi) - 2 \rho \Omega \, |\nabla_x \varphi|^2  + \rho \Omega^\bot \Delta \varphi, \\
& & \hspace{-1cm} \Omega^\bot \cdot \Delta (\rho \Omega) = \rho \Delta \varphi + 2  \, (\nabla_x \rho \cdot \nabla_x \varphi). 
\end{eqnarray*}
Therefore, system (\ref{Gene_mass}), (\ref{Gene_momentum}) is written:
\begin{eqnarray}
&& \hspace{-1cm} (\partial_t + \Omega \cdot \nabla_x) \rho + \rho \, (\Omega^\bot \cdot \nabla_x) \varphi = 0 , \label{2D_mass} \\
&& \hspace{-1cm} (\partial_t + c \Omega \cdot \nabla_x) \varphi + \frac{p'(\rho)}{\rho} \,  (\Omega^\bot \cdot \nabla_x) \rho = \mu \left( \Delta \varphi + 2 \frac{\nabla_x \rho \cdot \nabla_x \varphi}{\rho} \right) . \label{2D_momentum}
\end{eqnarray}
Introduce $\hat \rho = a(\rho)$ and $\lambda(\hat \rho)$ such that
\begin{equation} 
a'(\rho) = \frac{\sqrt{p'(\rho)}}{\rho}, \quad \lambda(\hat \rho) = a'(\rho) \rho, \quad h(\hat \rho)  = 2 \, \mbox{ln} \rho. 
\label{eq:def_a_lambda}
\end{equation}
Then, system (\ref{2D_mass}), (\ref{2D_momentum}) becomes: 
\begin{eqnarray}
&& \hspace{-1cm} (\partial_t + \Omega \cdot \nabla_x) \hat \rho + \lambda(\hat \rho) \, (\Omega^\bot \cdot \nabla_x) \varphi = 0 , \label{2DS_mass} \\
&& \hspace{-1cm} (\partial_t + c \Omega \cdot \nabla_x) \varphi + \lambda(\hat \rho) \,  (\Omega^\bot \cdot \nabla_x) \hat \rho = \mu \left( \Delta \varphi + \nabla_x h(\hat \rho) \cdot \nabla_x \varphi \right) . \label{2DS_momentum}
\end{eqnarray}
From (\ref{2DS_mass}), we have the following a priori estimate (maximum principle): 
\begin{equation} \rho_{\mbox{{\scriptsize min}}} \, \exp ( - \int_0^t \|\nabla_x \varphi (\cdot , s) \|_{L^\infty(\Pi^2)} \, ds ) \leq \rho \leq \rho_{\mbox{{\scriptsize max}}} \, \exp ( \int_0^t \|\nabla_x \varphi (\cdot , s) \|_{L^\infty(\Pi^2)} \, ds ) , \label{eq:max_principle}
\end{equation}
where
$$ \rho_{\mbox{{\scriptsize min}}} = \min_{x \in \Pi^2} \rho_0(x), \quad \rho_{\mbox{{\scriptsize max}}} = \max_{x \in \Pi^2} \rho_0(x). $$

We remind the following lemmas \cite{Taylor}: 

\begin{lemma}
For any pair of functions $f$, $g$ in $H^m({\mathbb R}^n) \cap L^\infty ({\mathbb R}^n) $, we have: 
\begin{eqnarray*}
& & \hspace{-1cm} \|fg\|_{H^m} \leq  C \left( \|f\|_{H^m} \|g\|_{L^\infty} + \|f\|_{L^\infty} \|g\|_{H^m} \right) .
\end{eqnarray*}
If additionally, we suppose that $\nabla f \in L^\infty ({\mathbb R}^n) $, we have, for any $\alpha \in {\mathbb N}^n$, with $|\alpha| = \sum_{i=1}^n \alpha_i = m$ :
\begin{eqnarray*}
& & \hspace{-1cm} \|D^\alpha (fg) - f D^\alpha g\|_{H^m} \leq C \left( \|f\|_{H^m} \|g\|_{L^\infty} +  \|\nabla_x f\|_{L^\infty} \|g\|_{H^{m-1}}  \right), 
\end{eqnarray*}
where $D^\alpha = \partial_{x_1^{\alpha_1} \ldots x_n^{\alpha_n}}$. 
\label{lem:Taylor}
\end{lemma}

Now, with $|\alpha| \leq m$, we take the $D^\alpha$ derivative of (\ref{2DS_mass}) and multiply it by $D^\alpha \hat \rho$ and integrate it with respect to $x$. Similarly, we take the $D^\alpha$ derivative of (\ref{2DS_momentum}) and multiply it by $D^\alpha \varphi$ and integrate it with respect to $x$. We sum up the resulting identities. Using the notation 
$$ \langle f,g \rangle = \int_{\Pi^2} f \, g \, dx, $$
we find:
\begin{eqnarray*}
 0 &=& \langle D^\alpha \hat \rho , D^\alpha \hat \rho_t \rangle +  \langle D^\alpha \varphi , D^\alpha \varphi_t \rangle   \\
&&+\langle D^\alpha \hat \rho , D^\alpha ((\Omega \cdot \nabla_x) \hat \rho) \rangle + c \langle D^\alpha \varphi , D^\alpha ((\Omega \cdot \nabla_x) \varphi) \rangle   \\
&&+\langle D^\alpha \hat \rho , D^\alpha (\lambda(\hat \rho) \, (\Omega^\bot \cdot \nabla_x) \varphi) \rangle +  \langle D^\alpha \varphi , D^\alpha (\lambda(\hat \rho) \, (\Omega^\bot \cdot \nabla_x) \hat \rho ) \rangle   \\
&&- \mu \langle D^\alpha \varphi , D^\alpha \Delta \varphi \rangle   \\
&&- \mu \langle D^\alpha \varphi , D^\alpha (\nabla_x h(\hat \rho) \cdot \nabla_x \varphi) \rangle \\
&=& I_1 + \ldots + I_5 
\end{eqnarray*}
Then: 
\begin{eqnarray*}
 I_1 &=& \frac{1}{2} \frac{d}{dt} ( \| D^\alpha \hat \rho \|^2 + \| D^\alpha \varphi \|^2 ) , 
\end{eqnarray*}
and 
\begin{eqnarray*}
 I_4 &=& \mu \| D^\alpha \nabla \varphi \|^2  ,
\end{eqnarray*}
where $\| \cdot \|$ just indicates an $L^2$ norm.  
Now, for the remaining terms, we have the following lemma

\begin{lemma}
We have: 
\begin{eqnarray*}
 |I_k| & \leq & C ( \| \hat \rho \|_{W^{1,\infty}} + \| \varphi \|_{W^{1,\infty}} ) \, ( \| \hat \rho \|_{H^m}^2 + \| \varphi \|_{H^m}^2 ) , \quad k=2, \, 3,  \\
 |I_5| & \leq & \frac{\mu}{2} \| \nabla D^\alpha \varphi \|^2 + 
 C ( \| \hat \rho \|_{W^{1,\infty}}^2 + \| \varphi \|_{W^{1,\infty}}^2 ) \, ( \| \hat \rho \|_{H^m}^2 + \| \varphi \|_{H^m}^2 ) , 
\end{eqnarray*}
where $C$ denote generic constants depending on the parameters of the problem. 
\label{lem:estimates}
\end{lemma}

The proof of the lemma is postponed at the end. 

Adding all these terms together for all possible indices $\alpha$ such that $|\alpha| \leq m$, we have, 
\begin{eqnarray*}
&& \frac{1}{2} \frac{d}{dt} ( \| \hat \rho \|_{H^m}^2 + \| \varphi \|_{H^m}^2  ) + \mu  \| \nabla \varphi \|_{H^m}^2 \leq \frac{\mu}{2} \| \nabla \varphi \|_{H^m}^2 + \\
& & \hspace{3cm} + C ( \| \hat \rho \|_{W^{1,\infty}}^2 + \| \varphi \|_{W^{1,\infty}}^2 + 1) \, ( \| \hat \rho \|_{H^m}^2 + \| \varphi \|_{H^m}^2  ) . 
\end{eqnarray*}
For $m \geq \frac{n}{2} + 1$, we have 
$$ \| \hat \rho \|_{W^{1,\infty}} + \| \varphi \|_{W^{1,\infty}} \leq 
C (\| \hat \rho \|_{H^m} + \| \varphi \|_{H^m} ) , $$
and get 
\begin{eqnarray*}
&& \frac{1}{2} \frac{d}{dt} ( \| \hat \rho \|_{H^m}^2 + \| \varphi \|_{H^m}^2  ) + \frac{\mu}{2} \| \nabla \varphi \|_{H^m}^2 \;\leq\; C \, ( \| \hat \rho \|_{H^m}^2 + \| \varphi \|_{H^m}^2  + 1)^2. 
\end{eqnarray*}
Gronwall's inequality leads to the local existence of a solution $(\hat \rho, \varphi)$ in $L^\infty([0,T],H^m(\Pi^2))$ which, if $\mu >0$, also belongs to $L^2([0,T],H^{m+1}(\Pi^2))$ and which satisfies the a priori bound (\ref{eq:max_principle}). To get time regularity, we directly use eqs. (\ref{2DS_mass}), (\ref{2DS_momentum}), take the $H^{m-1}$ norm, apply Lemma \ref{lem:Taylor}, and find 
$$ \| \hat \rho_t \|_{H^{m-1}} + \| \varphi_t \|_{H^{m-1}} \leq C \| \varphi \|_{H^{m+1}} + C ( \| \hat \rho \|_{W^{1,\infty}} + \| \varphi \|_{W^{1,\infty}} ) \, ( \| \hat \rho \|_{H^m} + \| \varphi \|_{H^m} ) . $$
Using the previous estimates, we deduce that $(\hat \rho, \varphi)$ also belongs to $H^1([0,T],H^{m-1}(\Pi^2))$. The estimates on $\hat \rho$ immediately transfer to $\rho$ since $a(\rho)$ is smooth and invertible for $\rho >0$. 
\endproof

\medskip
\noindent
{\bf Proof of Lemma \ref{lem:estimates}.} 
{\em Estimate of $I_5$:} Using Green's formula and Cauchy-Schwartz inequality, we have: 
\begin{eqnarray*}
|I_5| &\leq& \mu \| \nabla D^\alpha \varphi \| \, \| \nabla_x h(\hat \rho) \cdot \nabla_x \varphi \|_{H^{m-1}} \\
& \leq & C \| \nabla D^\alpha \varphi \| \, \left( \|\hat \rho \|_{H^m} \| \nabla \varphi\|_{L^\infty} + \|\nabla \hat \rho \|_{L^\infty} \|\varphi\|_{H^m} \right) \\
& \leq & \frac{\mu}{2} \| \nabla D^\alpha \varphi \|^2 + C \, \left( \|\hat \rho \|_{H^m} \| \nabla \varphi\|_{L^\infty} + \|\nabla \hat \rho \|_{L^\infty} \|\varphi\|_{H^m} \right)^2 \\
& \leq &  \frac{\mu}{2} \| \nabla D^\alpha \varphi \|^2 + 
 C ( \| \hat \rho \|_{W^{1,\infty}}^2 + \| \varphi \|_{W^{1,\infty}}^2 ) \, ( \| \hat \rho \|_{H^m}^2 + \| \varphi \|_{H^m}^2 ) . 
\end{eqnarray*}
The second inequality uses Lemma \ref{lem:Taylor} and the third one uses Young's inequality. \\

\noindent
{\em Estimate of $I_3$:} We write 
\begin{eqnarray*}
I_3 & = & \langle D^\alpha \hat \rho ,  \lambda(\hat \rho) \, (\Omega^\bot \cdot \nabla_x) D^\alpha \varphi \rangle \,+\,  \langle D^\alpha \varphi ,  (\lambda(\hat \rho) \, (\Omega^\bot \cdot \nabla_x) D^\alpha \hat \rho ) \rangle   \\
& & + \langle D^\alpha \hat \rho ,  \big( D^\alpha (\lambda(\hat \rho) \, (\Omega^\bot \cdot \nabla_x) \varphi) - \lambda(\hat \rho) \, (\Omega^\bot \cdot \nabla_x) D^\alpha \varphi\big)  \rangle \\
& & + \langle D^\alpha \varphi , \big(D^\alpha (\lambda(\hat \rho) \, (\Omega^\bot \cdot \nabla_x) \hat \rho ) - \lambda(\hat \rho) \, (\Omega^\bot \cdot \nabla_x) D^\alpha \hat \rho \big) \rangle \\
&=& J_1 + J_2 + J_3.
\end{eqnarray*}
Using Green's formula, we find 
\begin{eqnarray*}
|J_1| & = & |\langle \nabla \cdot (\lambda (\hat \rho) \, \Omega^\bot) \, D^\alpha \hat \rho ,  D^\alpha \varphi \rangle | \\
& \leq & C ( \| \hat \rho \|_{W^{1,\infty}} + \| \varphi \|_{W^{1,\infty}} ) \, ( \| \hat \rho \|_{H^m}^2 + \| \varphi \|_{H^m}^2 ) . 
\end{eqnarray*}
Now, using Cauchy-Schwartz inequality and applying Lemma \ref{lem:Taylor}, we find that $J_2$ and $J_3$ satisfy the same inequality.

\noindent
{\em Estimate of $I_2$:} The proof is similar as for $I_3$ and is omitted. \endproof

\subsection{Existence in 3D without viscosity}
\label{sub:existence_3D}

In this section, we investigate the local existence for the inviscid problem in 3 dimensions: 
\begin{eqnarray}
&& \hspace{-1cm} \partial_t \rho + \nabla_x \cdot (\rho \Omega) = 0 , \label{Gene3D_mass} \\
&& \hspace{-1cm} \rho (\partial_t \Omega + c \Omega \cdot \nabla_x \Omega ) + P_{\Omega^\bot} \nabla_x (p(\rho)) = 0 , \label{Gene3D_momentum}
\end{eqnarray}
where the parameters and data have the same meaning as in section \ref{sub:existence_2D}. We consider the system in the domain $\Pi^3 = [0,1]^3$ with periodic boundary conditions. 

For this purpose, we use the spherical coordinates associated to a fixed Cartesian basis. In this basis, denoting by $\theta \in [0,\pi]$ the latitude and $\varphi \in [0,2 \pi]$ the longitude, we have 
$$ \Omega = (\sin \theta \, \cos \varphi, \sin \theta \,  \sin \varphi, \cos \theta)^T, $$
and we let $\Omega_\theta$ and $\Omega_\varphi$ be the derivatives of $\Omega$ with respect to $\theta$ and $\varphi$. We note that 
$$ |\Omega_\theta|=1, \quad |\Omega_\varphi|= \sin \theta. $$
We will use the formulas
\begin{eqnarray*}
& & \hspace{-1cm} \nabla_x \cdot \Omega = \Omega_\theta \cdot \nabla_x \theta + \Omega_\varphi \cdot \nabla_x \varphi, \\
& & \hspace{-1cm} P_{\Omega^\bot} a =  (\Omega_\theta \cdot a) \Omega_\theta + \frac{(\Omega_\varphi \cdot a)}{\sin^2 \theta} \Omega_\varphi , \\
& & \hspace{-1cm} (\Omega \cdot \nabla_x) \Omega = \big((\Omega \cdot \nabla_x)\theta\big) \, \Omega_\theta + \big((\Omega \cdot \nabla_x)\varphi\big) \, \Omega_\varphi, \\
& & \hspace{-1cm} \Omega_t = \Omega_\theta \, \theta_t + \Omega_\varphi \, \varphi_t , 
\end{eqnarray*}
where $a$ is an arbitrary vector. 

Introduce $\hat \rho$ and $\lambda(\hat \rho)$ as in (\ref{eq:def_a_lambda}). Then, system (\ref{Gene3D_mass}), (\ref{Gene3D_momentum}) becomes:
\begin{eqnarray*}
&& \hspace{-1cm} \hat \rho_t + \Omega \cdot  \nabla_x \hat \rho + \lambda(\hat \rho)  \nabla_x \cdot \Omega = 0, \\
&& \hspace{-1cm} \Omega_t + c \,  (\Omega \cdot \nabla_x) \Omega + \lambda(\hat \rho)  \, P_{\Omega^\bot} \nabla_x \hat \rho = 0 , \end{eqnarray*}
or, 
\begin{eqnarray}
&& \hspace{-1cm} \hat \rho_t + \Omega \cdot  \nabla_x \hat \rho + \lambda(\hat \rho)  (\Omega_\theta \cdot \nabla_x \theta + \Omega_\varphi \cdot \nabla_x \varphi) = 0, \label{eqnv:rho_limit} \\
&& \hspace{-1cm} \theta_t + c \,  (\Omega \cdot \nabla_x) \theta   + \lambda(\hat \rho) \, \Omega_\theta \cdot \nabla_x \hat \rho = 0 , \label{eqnv:theta_limit} \\
&& \hspace{-1cm} \sin^2 \theta \varphi_t + c \, \sin^2 \theta \,  (\Omega \cdot \nabla_x) \varphi   + \lambda(\hat \rho) \, \Omega_\varphi \cdot \nabla_x \hat \rho = 0 . \label{eqnv:varphi_limit} 
\end{eqnarray}
Introducing
$$ U = \left( \begin{array}{c} \hat \rho \\ \theta \\ \varphi \end{array} \right), $$
this system is written 
$$ A_0(U) U_t + A_1(U) U_x + A_2(U) U_y + A_3(U) U_z = 0, $$
in Cartesian coordinates $x = (x,y,z)$, where $A_k(U)$, $k=0, \ldots , 4$ are all symmetric matrices and 
$$ A_0 = \left( \begin{array}{ccc} 1 & 0 & 0 \\ 0 & 1 & 0 \\ 0 & 0 & \sin^2 \theta \end{array} \right) . $$
If $\sin \theta > 0$, then this system is a symmetrizable hyperbolic system. We can apply proposition 2.1 p. 425 of \cite{Taylor} and the following theorem follows immediately:

\begin{theorem}
We assume that the initial data $(\rho_0, \theta_0, \varphi_0)$ belong to $H^m(\Pi^3)$ with $m>5/2$ with $\rho_0 >0$, $\sin \theta_0 >0$. Then, there exists a time $T>0$ and a unique solution $(\rho, \theta, \varphi)$ in $L^\infty([0,T],H^m(\Pi^3)) \cap H^1([0,T],H^{m-1}(\Pi^3))$ such that $\rho$ remains positive. 
\label{thm:exist_3D}
\end{theorem}

\setcounter{equation}{0}
\section{Conclusion}
\label{sec:conclu}

In this paper, we have derived hydrodynamic systems from kinetic models of self-propelled particles with alignment interaction and attraction-repulsion force. We have particularly focused on the inclusion of diffusion terms under the assumption of weakly non-local interactions. Then, we have proved the local-in-time existence of solutions for the viscous system in 2D and a similar result for the inviscid system in 3D. The methods rely on a suitable symmetrization and on the energy method. Future works in this direction will consist in continuing the exploration of the mathematical structure of the system and particularly, trying to prove local existence of the viscous system in 3D and the treatment of the geometric singularity near $\sin \theta = 0$. Another direction of work will consist of the numerical quantification of the viscosity as a consequence of the non-locality of the interaction.


\end{document}